# Fabrication and characterization of an undoped GaAs/AlGaAs quantum dot device

Hai-Ou Li,[1,2] Gang Cao,[1,2] Ming Xiao,[1,2,a] Jie You,[1,2] Da Wei,[1,2]
Tao Tu,[1,2] Guang-Can Guo,[1,2] Hong-Wen Jiang,[3] and Guo-Ping Guo[1,2,b]

[1] *Key Laboratory of Quantum Information, CAS, University of Science and Technology of China, Hefei, Anhui 230026, China*

[2] *Synergetic Innovation Center of Quantum Information & Quantum Physics, University of Science and Technology of China, Hefei, Anhui 230026, China*

[3] *Department of Physics and Astronomy, University of California, Los Angeles, CA 90095, USA*

a),b) *Authors to whom correspondence should be addressed. Electronic mails: maaxiao@ustc.edu.cn; gpguo@ustc.edu.cn*

**ABSTRACT**

We demonstrate the development of a double quantum dot with an integrated charge sensor fabricated in undoped GaAs/AlGaAs heterostructures using a double top-gated design. Based on the evaluation of the integrated charge sensor, the double quantum dot can be tuned to a few-electron region. Additionally, the inter-dot coupling of the double quantum dot can be tuned to a large extent according to the voltage on the middle gate. The quantum dot is shown to be tunable from a single dot to a well-isolated double dot. To assess the stability of such design, the potential fluctuation induced by $1/f$ noise was measured. Based on the findings herein, the quantum dot design developed in the undoped GaAs/AlGaAs semiconductor shows potential for the future exploitation of nano-devices.

Electrically gate-defined semiconductor quantum dots fabricated on modulation doped GaAs/AlGaAs heterostructures are widely used to explore electron transport behaviors and to realize solid state qubits.[1-8] They have attracted much interest because of their remarkable coherence controllability by tuning parameters in a two-level system with electrical gates.[9-13] However, it is believed that doped heterostructures may suffer from significant charge noise owing to surface gate leakage and charge fluctuation in the doping layer,[14-18] thereby resulting in an important decoherence source for solid state qubits.[15,16] Removal of the ionized doping layer, leading to an undoped semiconductor, may provide a good platform for implementing various quantum devices. Because an additional top gate is used to accumulate charge carriers under the conventional depletion gates, the double-gate architecture is usually adopted in undoped semiconductor devices. Recently, there have been a series of reports on double-gated quantum dots in undoped Si/SiO$_2$[19-21] and Si/SiGe materials.[22-24]

Undoped GaAs/AlGaAs heterostructures have better interfacial properties than Si/SiO$_2$[19] and have no valley interference as observed for Si/SiGe.[24] By employing undoped GaAs/AlGaAs systems, researchers have successfully developed single quantum dots with electron carriers[25,26] and double quantum dots with hole carriers.[27] Double quantum dots with electron carriers are expected to be useful for further investigation of the coherence controllability and decoherence noise level of single electron-based quantum devices in undoped GaAs/AlGaAs.

In this letter, we report an n-type double quantum dot (DQD) device in an undoped GaAs/AlGaAs heterostructure. An integrated quantum point contact (QPC) acts as a charge sensor. The quantum dot system can be tuned from a single dot to a well-isolated double dot in a few-electron region, and the inter-dot coupling between the left and right dots can be largely tuned according to the voltage on the middle gate. By monitoring the 1/$f$ noise level around a Coulomb blockade peak, a corresponding potential fluctuation in the order of 1 μeV is obtained, similarly to results in shallow-etched GaAs quantum dot.[28] The high tunability of tunnel coupling and lower levels of charge noise present potential for the development of various quantum devices in undoped GaAa/AlGaAs heterostructures.

The current quantum dot design was based on a double top-gated architecture, similar to previous work conducted on GaAs[27] and Si.[21] A schematic illustration of the cross-section of the device is shown in Fig. 1(a); the device structure was formed by molecular beam epitaxy on strain-relaxed Al$_{0.3}$Ga$_{0.7}$As buffers. The global gate consisting of a 200-nm-thick layer of thermally evaporated Al was used to accumulate a two-dimensional electron gas (2DEG) at the interface between GaAs and Al$_{0.3}$Ga$_{0.7}$As, 45 nm below the heterostructure surface. A set of local depletion gates patterned with Ti and Au defined a confining potential to form the quantum dots. Ohmic contacts to the electron layer were formed via Ni/AuGe evaporation and annealing, as described in ref.[29]: Ni(10 nm)/AuGe(150 nm) was evaporated at an angle of 60° relative to the normal plane to ensure good side-wall wetting, and then annealed at 430 ℃ for 30 min in an N$_2$/H$_2$ atmosphere. A 100-nm-thick layer of Al$_2$O$_3$ grown via atomic layer deposition electrically isolated the global gate from the depletion gates and Ohmic contacts.

Figure 1(b) shows a scanning electron image of a typical set of depletion gates, defining a double quantum dot. Gates T–L and T–R form tunnel barriers and gate T–M controls the inter-dot coupling between the left and right dots. Gates LP and RP are used to adjust the energy level of each dot. Combination of gates Q1 and Q2 defines a QPC, which is capacitively coupled to the dots and is able to measure the charge occupation of the device. Ohmic connections to 2DEGs are represented by white boxes. Current through the dot is measured between S and D, and QPC current is measured between A and S.

The experiment was performed in a He3 refrigerator at the base temperature of 240 mK. For the quantum dot transport and QPC charge counting measurements, the standard lock-in technique was used. The noise spectra was obtained through the fluctuation of the QD current using a fast network analyzer SR785. First, a van der Pauw sample was prepared (whereby the depletion gates were omitted while other parameters were retained as per the set-up in Fig. 1(a)) and the quantum hall effects for bulk 2DEG were assessed. Figure 1(c) shows a plot of the electron density $n_{2DEG}$ as a function of global top gate bias voltage $V_{Top}$. The red circles represent the experiment data and the dashed line represents the linear fit curve of the data, yielding a capacitance of $3.2 \times 10^{11}$ cm$^2$ V$^{-1}$. Figure 1(d) shows a plot of the measured electron mobility $\mu_{2DEG}$ as a function of $n_{2DEG}$. For all the latter measurements, the global gate is chosen so that $n_{2DEG}$ is in the range of (2–3) $\times 10^{11}$ cm$^{-2}$ and $\mu_{2DEG}$ correspondingly lies in the range of (1.5–2.0) $\times 10^5$ cm$^2$ V s$^{-1}$.

When the voltages at gates T and M were significantly positive, the inter-dot coupling was considerably strong, allowing the device to operate as a large single dot. Figure 1(e) shows the dot transport conductance as a function of $V_{LP}$ and $V_{RP}$ under the following conditions: $V_T = 0.16$ V and $V_M = 0.6$ V. The single set of parallel lines confirms that the device is composed of a large single dot. Figure 1(f) shows the stability diagram for the single dot. The total capacitance between the dot and all gates was $C = 113$ aF (charging energy $E_C = e^2/C = 1.5$ meV) and the capacitance between the dot and gate RP was $C_{RP} = 3.1$ aF. This gives a lever arm $\alpha_{RP} = C_{RP}/C = 0.027$.

The charge-sensing measurements in Fig. 2(a)–(d) show the QPC transconductance $dI_{QPC}/dV_R$ as a function of the left and right barrier gate voltages, $V_L$ and $V_R$, at four different middle gate voltages $V_T = 0.44, 0.36, 0.32$, and $0.24$ V. Comparison of these data demonstrates the ability to use the middle gate T to tune the DQD from a strongly coupled regime (comparable to a large single dot) to a weakly coupled regime (behaving as two well-isolated dots). In Fig. 2(d), we noticed the absence of charge transition lines in the lower-left region of the plot across a wide voltage range. This suggests that the DQD is empty with zero-charge occupancy. The electron occupation ($M$, $N$) is shown in Fig. 2(d), where $M$ and $N$ indicate the number of electrons in the left and right dots, respectively.

Figure 3(a) shows the charge stability diagram in a weakly coupled region. The QPC modulation signal for inter-dot charge transitions, which represent a single-electron tunneling from one dot to the other, can be obtained across the dashed line. Evaluation of this transition affords the determination of numerous important quantities describing the double dot. DiCarlo et al.[30] first employed the extent of precision of such a conductance shift to investigate the effective electron temperature and coupling strength between dots. In our case, the QPC modulation signal, which is the physical derivative of the QPC conductance, was measured because of its apparent considerably better signal-to-noise ratios. Wei et al.[31] refined DiCarlo's technique for applicability towards a QPC modulation signal. Accordingly, the reported methods were used herein to determine three parameters: electron temperature $T_e$, lever arm $\alpha$, and coupling strength $t_C$.

$T_e$ can be deduced from the peak width of the QPC modulation signal, shown as a function of the cryostat temperature in Fig. 3(b). The result suggests that $T_e$ is equivalent to the lattice temperature (corresponding to the cryostat temperature) throughout the measurement. Additionally, $T_e$ under the cryostat base temperature of 240 mK was determined as ~300 mK. The lever arm of the left barrier gate, $\alpha_L = C_L/C_{L\text{-dot}} = 0.047$, was determined from the slope of the QPC modulation signal. Integration of the QPC modulation signal generates the QPC conductance; plots of the left-dot occupancy as a function of the left–right dot detuning $\varepsilon$ at varying $V_T$s within the weakly coupled region are shown in Fig. 3(c). At higher positive $V_T$s, the increased tunnel coupling $t_C$ results in clear broadening of the inter-dot charge transition line. Figure 3(d) shows a plot of $t_C$ as a function of $V_T$, using DiCarlo's fitting model based on the extracted parameters $T_e$ and $\alpha_L$. The dependence of $t_C$ on $V_T$ appears to be exponential. This clearly demonstrates that the inter-dot coupling of the double quantum dot can be tuned (from 25 to 120 μeV) by varying the voltage of the middle gate. Owing to limitations of the cryostat temperature, $t_C$ values smaller than 25 μeV could not be measured though we believe that it can be achieved.

To assess the stability of our device when adjusted as a single quantum dot, the $1/f$ noise around a coulomb blockade (CB) region peak was measured. Figure 4(a) shows a typical current–noise spectrum, $S_I(f)$, measured at three different regions A, B, and C in the inset. The inset is a plot of the Coulomb peak measured as a function of sweeping voltage $V_{RP}$ while $V_{SD}$ is biased at 200 μV. The spectrum at point A was measured in the CB region, and represents the noise level of the present measurement system. The spectra at points B and C show a $1/f$-like property at low frequencies. However, the spectrum at point C shows a smaller noise than at point B though the average current at point C is higher than that at point B. This indicates that the fluctuation of the tunneling current at point B is larger than that at point C. An explanation is given in the following paragraphs. Current fluctuation was estimated by integrating the spectrum within a limited frequency range of 5–45 Hz, as described in ref.[28]:

$$\Delta I = \sqrt{\int_5^{45}[S_I^2(f) - S_{CB}^2(f)]df} \quad (1)$$

The background noise spectrum $S_{CB}(f)$ is obtained in the CB region (point A) and $S_I(f)$ is the current noise spectrum measured at different points on the Coulomb peak (i.e., points B and C) ground states. Finally, the influence of the current slope change is subtracted to yield fluctuation of the potential in terms of energy, using the equation given in ref.[28]:

$$\Delta I = \alpha^{-1}|dI/dV|\Delta\varepsilon. \tag{2}$$

Herein, lever arm $\alpha_{RP} = C_{RP}/C$ was determined as 0.027; $dI/dV$ is the derivative of the current with respect to $V_{RP}$. The relationship between $\Delta I$ and $\Delta\varepsilon$ can be understood as a normalization process of the noise measured along the Coulomb peak; $\Delta\varepsilon$ can be regarded as a reasonable parameter for describing the overall noise level of quantum devices.

Figure 4(b) and (c) show the current fluctuation $\Delta I$ and potential fluctuation $\Delta\varepsilon$, respectively. As noted in Fig. 4(b), $\Delta I$ decreased to zero around point C. This is understandable according to Eq. (2): $dI/dV$ reaches zero at point C and in principle $\Delta\varepsilon$ is independent of the current level. However, herein, calculations of $\Delta\varepsilon$ were based on $\Delta I$. Therefore, the data points around point C in Fig. 4(c) become unacceptably large because of a nearly zero $|dI/dV|$. Following removal of these inaccurate data points, the potential fluctuation was measured in the range of 0.55–1.4 μeV, which is similar to report involving shallow-etched GaAs quantum dot.[28]

In summary, we reported the fabrication of double quantum dot devices in an undoped GaAs/AlGaAs heterostructure using a double top-gated design. The device can be adjusted to operate as a single dot or a well-isolated double dot. Charge counting signal shows that we can reliably deplete the DQD to zero-charge occupancy and operate the DQD in the few-electron regime. The coupling strength between the quantum dots as a function of the voltage on the middle gates was determined. The potential fluctuation induced by $1/f$ noise was measured. The current study is a gateway to future research on quantum devices based on undoped GaAs/AlGaAs heterostructures such as the coherent control of a single-electron charge on a DQD.

This work was supported by the National Fundamental Research Program (Grant No. 2011CBA00200), the National Natural Science Foundation (Grant Nos. 11222438, 61306150, 11174267, 11304301, 11274294, and 91121014), and the Chinese Academy of Sciences.

**Figure captions**

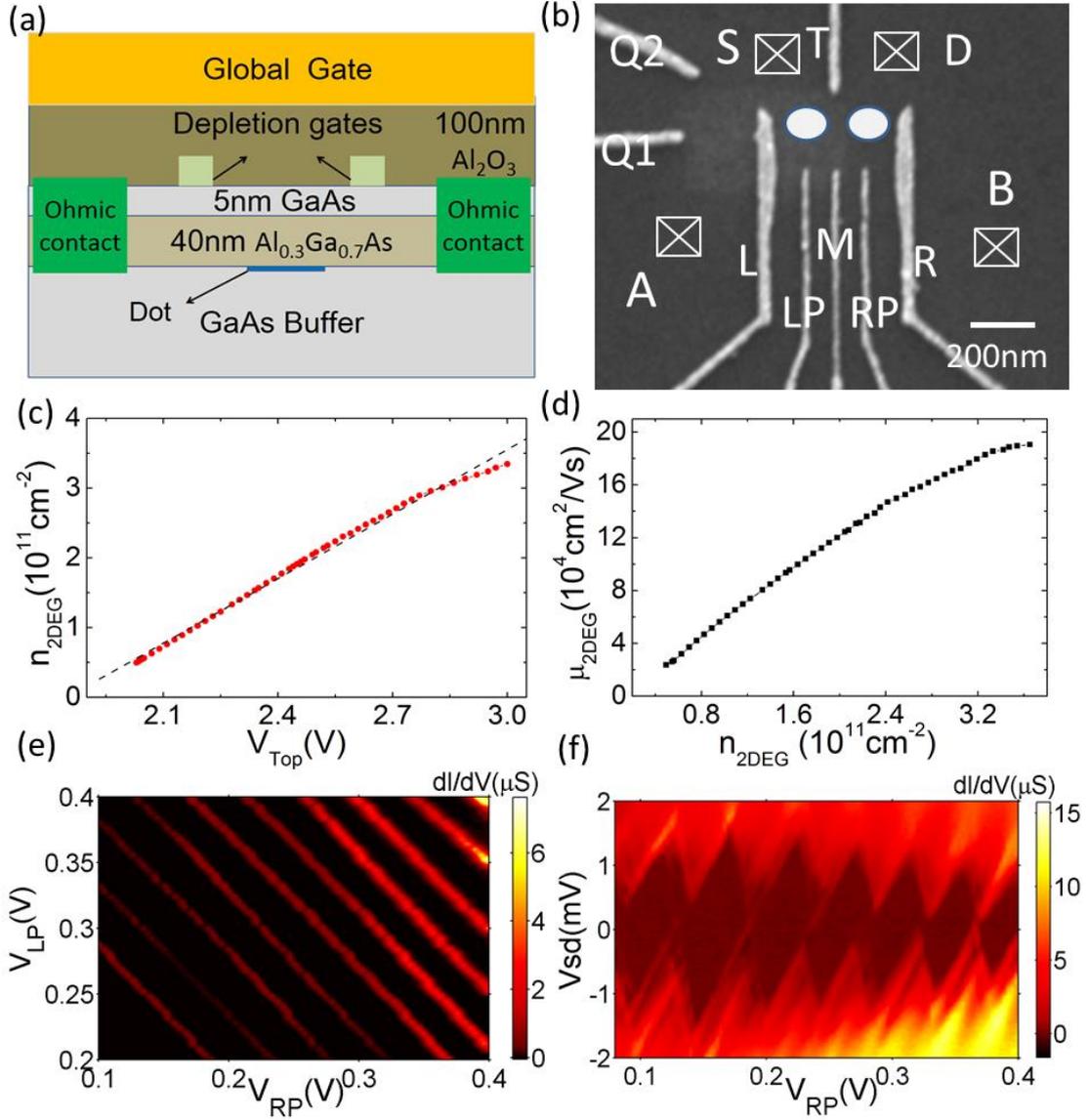

FIG. 1. (a) Schematic cross-section of the device. The DQD is operated by biasing a global gate at voltage $V_{Top}$ to accumulate carriers in the quantum well, and the local depletion gates define the DQD confinement potential. (b) Scanning electron image of the DQD device. Ohmic connections to 2DEGs are represented by white boxes. Current through the dot is measured between S and D, and the QPC current is measured between ohmic contacts A and S. (c) Plot of electron density $n_{2DEG}$ as a function of the global top gate bias $V_{Top}$ for bulk 2D devices; red circles represent experimental data and the dashed line represents the linear fit curve of the data. (d) Plots of electron mobility $\mu_{2DEG}$ as a function of $n_{2DEG}$. (e) Dot conductance measured in transport as a function of $V_{LP}$ and $V_{RP}$ for large single quantum dot with $V_{Top}$ = 3.2 V, $V_T$ = 0.16 V, $V_M$ = 0.6 V, and $V_L = V_R$ = 0.3 V. (f) Coulomb diamonds measured in transport through the quantum dot operating as a single quantum dot with $V_{Top}$ = 3.2 V, $V_T$ = 0.16 V, $V_M$ = 0.6 V, $V_L = V_R$ = 0.32 V, and $V_{LP}$ = 0.3 V.

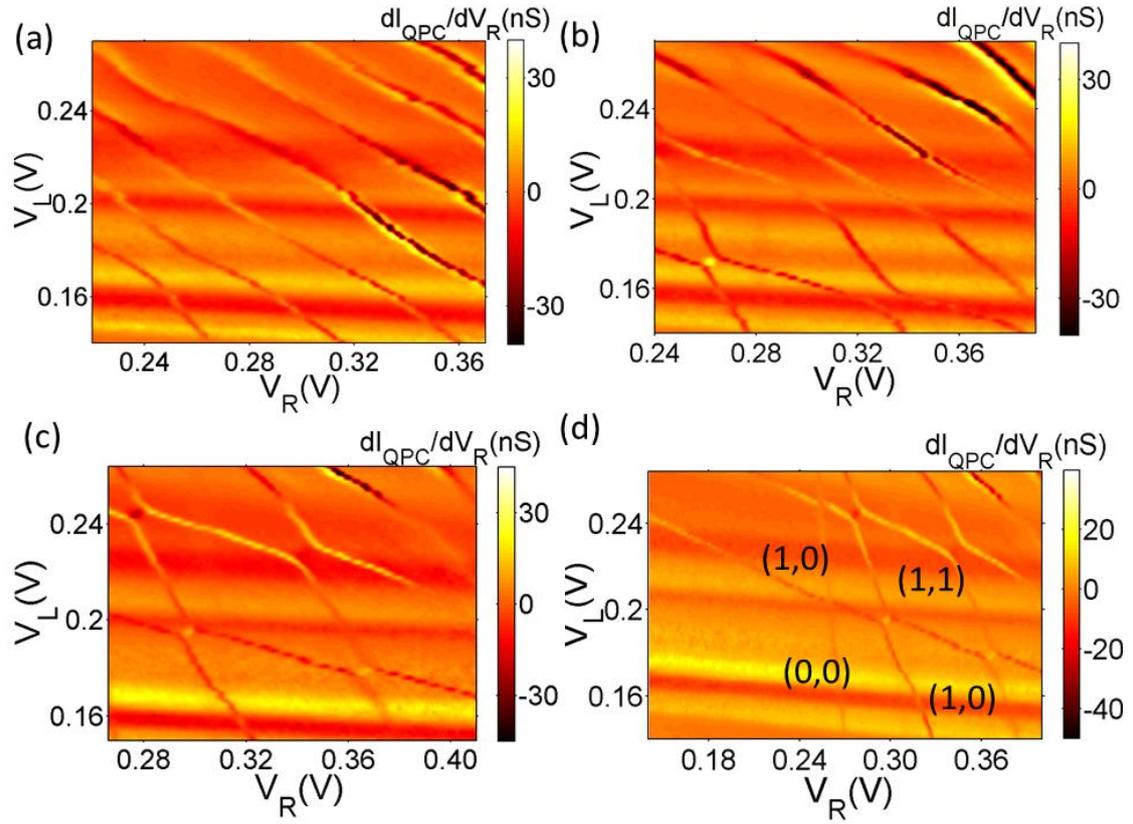

FIG. 2. Representative QPC charge sensor stability diagrams of the double quantum dot. Images showing the QPC transconductance d$I_{QPC}$/d$V_R$ as a function of gate voltages $V_L$ and $V_R$ at different middle gate voltages $V_T$ of (a) 0.44, (b) 0.36, (c) 0.32, and (d) 0.24 V (the DQD is empty with zero-charge occupancy); $V_{Top}$ = 3.2 V, $V_{LP}$ = 0.2 V, $V_{RP}$ = 0.2 V, and $V_M$ = −0.5 V.

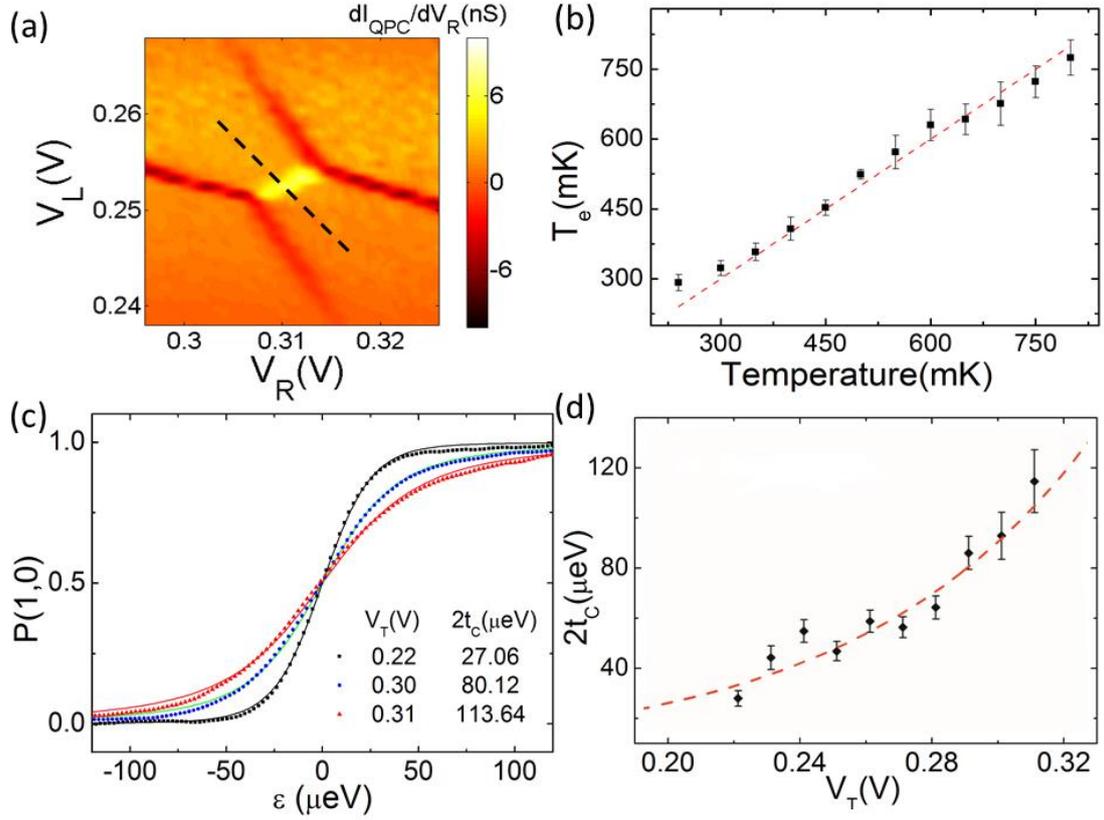

FIG. 3. Electron temperature and tunnel coupling tuning measurements. (a) Charge stability diagram in a weakly coupled region with $V_T = 0.22$ V; the black dashed line indicates the scanning direction for the transition peak. (b) Plot of electron temperature as a function of temperature. The observed positive linearity suggests that the electron temperature is equal to the lattice temperature throughout the measurements. (c) Plots of P(1,0) as a function of detuning at different $V_T$ values; plots show tunable inter-dot tunnel coupling at the (1,0)–(0,1) inter-dot charge transition. (d) Plot of inter-dot tunnel coupling as a function of the middle gate voltage $V_T$. Results show that $2t_C$ can be tuned and increased from 25 to 120 μeV, in an exponential fashion, with increasing gate voltages.

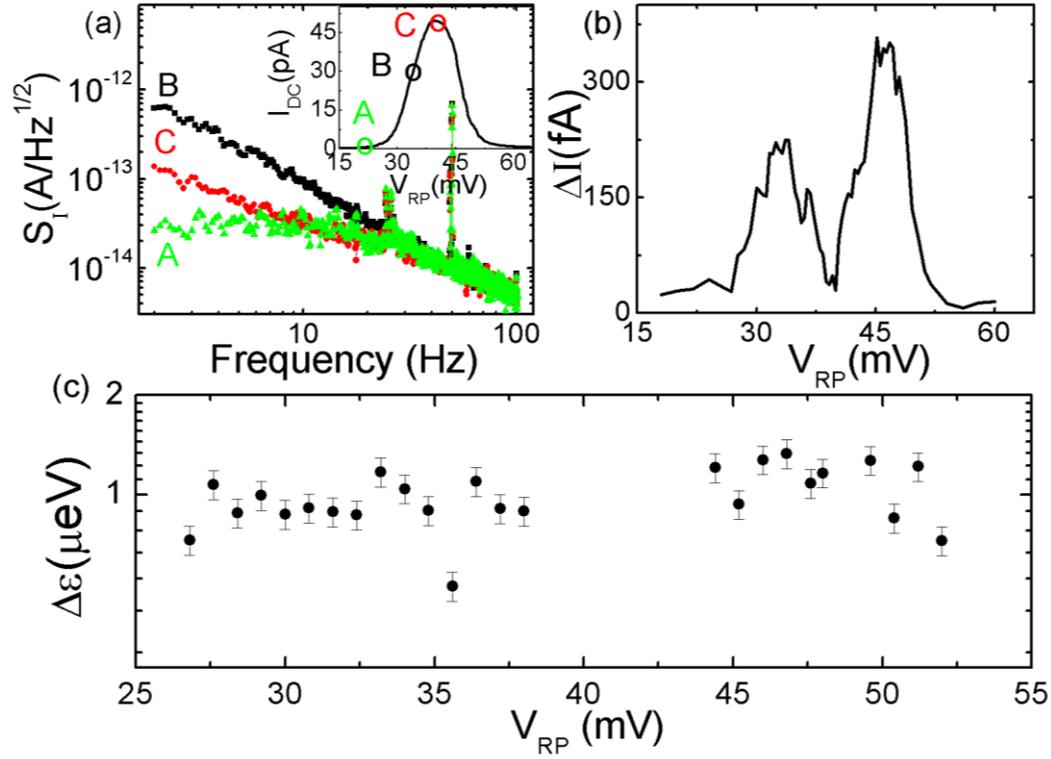

FIG. 4. (a) Noise spectra measured at different regions of the Coulomb peak, denoted as A, B, and C in the inset; the Coulomb peak is defined as the typical tunneling current peak measured when depletion gate voltage $V_{RP}$ is swept at $V_{SD} = 200$ μV. (b) Plot of current fluctuation $\Delta I$ integrated from 5 to 45 Hz of the peak shown in (a) as a function of depletion gate voltage $V_{RP}$. (c) Plot of potential fluctuation $\Delta\varepsilon$ as a function of $V_{RP}$.